# Leaf Water Status Monitoring by Scattering Effects at Terahertz Frequencies

Bin Li[1], Xiao Zhang[2], Rong Wang[2], Yu Mei[2], and Jianjun Ma[2]

[1]Beijing Research Center for Information Technology in Agriculture, Beijing 100097, China

[2]School of Information and Electronics, Beijing Institute of Technology, Beijing 100081, China

* Correspondence: jianjun_ma@bit.edu.cn

**Abstract**

The applications of terahertz (THz) radiation for plant water status monitoring require systematic studies on interaction of THz wave and plants. Here, we present theoretical investigations on scattering behavior of THz waves reflected by and transmitting through a plant leaf under different water content. A theoretical model combining integral equation and radiative transfer theory is presented to fit the measured data. Good agreement confirms the availability of the model for water status evaluation when variation of leaf thickness and surface roughness is considered. We investigate the applicability of THz waves for water status monitoring in reflection and transmission geometries under different temperatures, salinities and polarizations.

**Keywords:** Terahertz wave; water absorption; water status monitoring; reflection; transmission; integral equation; radiative transfer theory

## 1. Introduction

Water status of plants is usually regarded as an implication of their different exogenous and physiological conditions, which could affect productivity of crops [1]. Water status variation is regarded as an important indicator of crop health to farmers, agricultural scientists and plant physiologists. It provides valuable information in irrigation management, helps to avoid irreversible damage, and thus substantially reduces or prevents yield losses [2]. So, it is definitely necessary to develop methods which can assess water content of different types of crops.

Present techniques for plant water status evaluation can be divided into two groups, destructive [3, 4] and non-destructive [5] methods. The destructive method offers high reliability and ease of handling. But it leads to irrevocable damage and is not preferred for long-term researches of the same plant. However, the non-destructive technique can do measurements without any contact with plants, which is always required over a long duration monitoring, even though its accuracy limitation always exists and imprisons its applications.

THz monitoring is found to be a promising technique for water content assessment with numerous advantages, such as superior spatial resolution, high transparency through many common materials and non-ionizing [6-9]. This method employs the correlation between water status and attenuation of transmitted THz signal through the sample due to its high sensitivity to water molecule. This makes it to be a sensitive and precise tool for water status monitoring [10-12] and product quality control [13, 14].

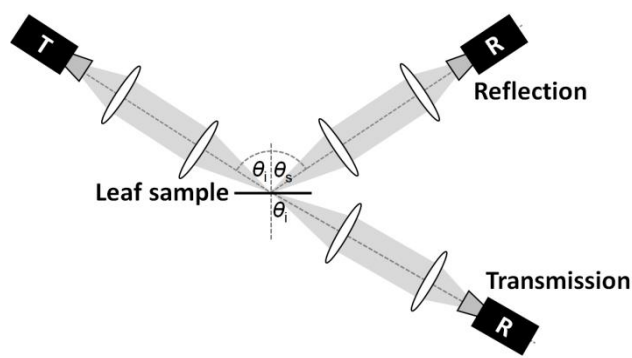

Figure 1 Basic reflection and transmission geometries for THz monitoring technique.

However, this approach, usually operating in transmission geometry (Fig. 1), is limited significantly when water absorption is strong. Reflection geometry could become a more efficient method for thick plant samples [15, 16], which could reduce the effects of absorption by water inside. Fortunately, the thickness of most plant leaves, such as ginkgo [3], coffee [17], barley [7] and arabidopsis thaliana [18], is very thin and both the transmission and reflection geometries could be applied.

Despite the appeal of such applications, the ability and limitation of THz radiation in leaf water status monitoring still needs systematic studies because of the difficulties in leaf characterization and experimental measurement due to the complicated influencing factors, such as temperature, salinity, humidity and so on. Therefore, fundamental studies on the interaction of THz interaction with plant leaves and modeling of THz performance under realistic conditions are valuable. Ray-tracing and propagation matrix were employed to describe THz transmission through single and multiple layers of coffee leaves [19], but surface scattering due to leaf roughness is not included, which could affects the reflection and transmission seriously [20]. A Rayleigh factor is introduced in [17] and better agreement between measurement and simulation is observed. However, it just fits to specular reflection and neglects the surface roughness distribution which usually follows several kinds of correlation functions. Present methods are not enough to describe the scattering behavior by plant leaves with high surface roughness [21]. In this paper, we try to investigate the influence of surface and volume scattering on the plant leaf water status assessment, and develop a theoretical model including temperature variation, salinity and wave polarization in both reflection and transmission geometries.

## 2. Physical Characteristics of Plant Leaves

A description of the interaction between the THz radiation and a plant leaf requires detailed structural information of the leaf, which consists of a complex biological structure of tissues and different biomolecules, such as cellulose and other carbohydrates, proteins, and many other high and low molecule weight compounds [17]. But here, it is not necessary to consider all the components because of their much smaller size compared with THz wavelength in the range of mm and sub-mm. A leaf can be considered as a compound of a non-dispersive residual component and highly dispersive water [22]. The water is distributed continuously inside the leaf without any shape, so the effective medium theory, such as Maxwell-Garnett, Polder and van Santen, Extended Bruggeman, Landau-Lifshitz-Looyenga (LLL) and Complex Refractive Index, would not be considered, because of the assumption of spherical, ellipsoidal or arbitrarily shaped inclusions [7, 23-26]. The Debye-Cole model is one of the most widely used models for estimation of dielectric properties of plant leaves. It treats the vegetation as a mixture of bulk vegetation, free water component and bound water component [27] and can be expressed as

$$\varepsilon = \varepsilon_r + V_{fw}\varepsilon_w + V_{bw}\varepsilon_b \tag{1}$$

where, $\varepsilon_r = 1.7 - 0.74 m_g + 6.16 m_g^2$ is the dielectric constant of the non-dispersive residual component. $m_g$ is gravimetric water content and can be obtained by $m_g = (m_{fresh} - m_{dry})/m_{fresh} \times 100\%$ with $m_{fresh}$ and $m_{dry}$ as the mass of fresh and dry leaves. The including of water content in above equation shows better fit to measurements. $\varepsilon_r$ varies from 0.7 for dry leaf to 4.5 for a gravimetric water content of 0.7. $\varepsilon_w$ and $\varepsilon_b$ are the complex dielectric constants of free water and bound water, respectively, with $v_{fw} = 0.55 m_g^2 - 0.076 m_g$ and $v_{bw} = 4.64 m_g^2 / (1 + 7.36 m_g^2)$ as their associated volume fractions. Fig. 2 shows the variation of the fractions with respect to water content $m_g$. It can be seen that the bound water is always a major part of water in leaf.

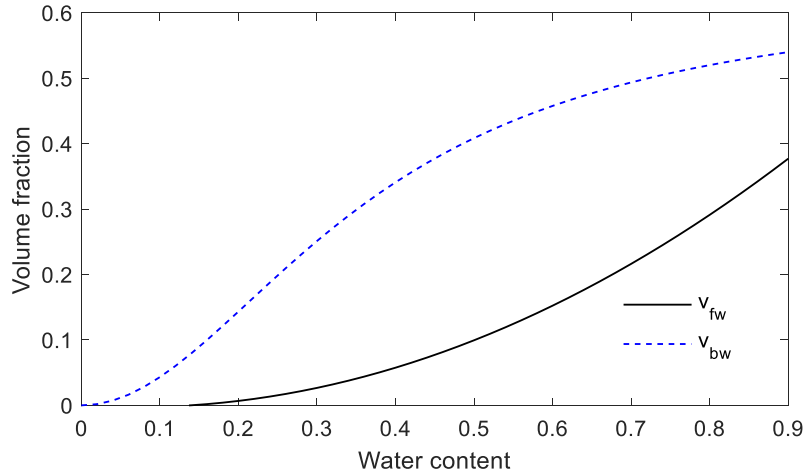

Figure 2 Variation of volume fractions of free water and bound water with respect to gravimetric water content $m_g$.

Double-Debye dielectric model (D3M) was found to be the most accurate model to estimate the dielectric constant of water and was developed for saline water [28-30] as

$$\varepsilon_w = \varepsilon_{w\infty} + \frac{\varepsilon_{w0} - \varepsilon_{w\infty}}{1 + (f/f_0)^2} - j\left[\frac{(f/f_0)(\varepsilon_{w0} - \varepsilon_{w\infty})}{1 + (f/f_0)^2} + \frac{\sigma_i}{2\pi\varepsilon_0 f}\right] \quad (2)$$

with $f_0$ as its relaxation frequency, which is related to temperature. $\varepsilon_{w0}$ is the static dielectric constant when $f = 0$, $\varepsilon_{w\infty}$ is the high frequency dielectric constant when $f \to \infty$ and $\varepsilon_0$ is the permittivity of free space. $\sigma_i$ is ionic conductivity of the saline water affected by both temperature $T$ and salinity $S$ as $\sigma_i = \sigma(T, 35) \cdot P(S) \cdot Q(T, S)$, because the free water in vegetation includes low concentration of salts and sugars. Expressions for the parameters $\sigma(T, 35)$, $P(S)$ and $Q(T, S)$ are presented in [28, 31]. The salinity was introduced to improve the accuracy of the model because the free water in vegetation includes low concentration of salts and sugars [32], which would affect the complex permittivity. Expressions for the parameters could be obtained in [28, 33] and the Eq. (2) is plotted in Fig. 3 under different salinity and temperatures. The dielectric constant of free water is sensitive to salinity and temperature, which is proved in [34, 35].

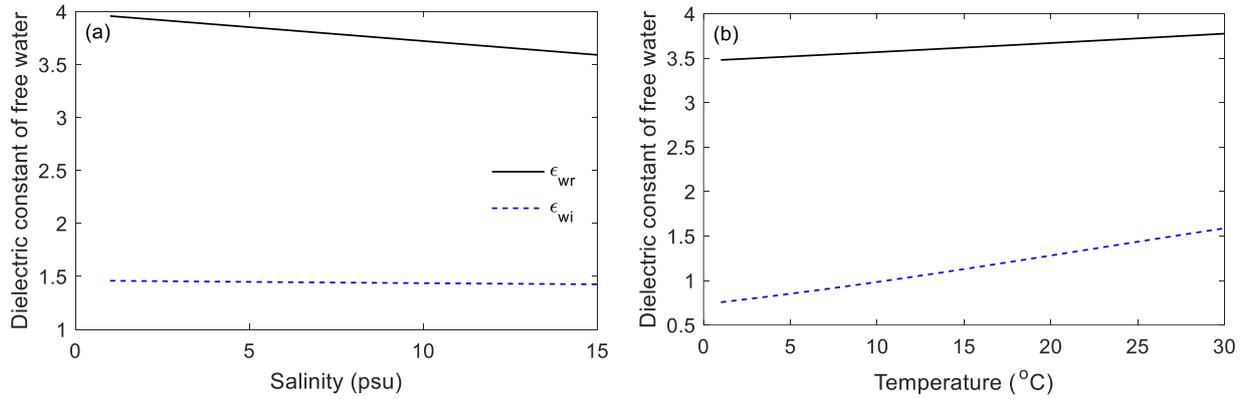

Figure 3 Variation of dielectric constant of free water with respect to (a) salinity and (b) temperature with THz frequency at 1 THz, water content $m_g = 0.5$, salinity $S= 10$ psu and temperature $T= 25^\circ$C.

For the bound water component, the relaxation time $\tau$ of a water molecule is longer than that in the free state, because it is governed by the interaction of the water molecule and its environment. In bound state, the molecule is under the influence of non-electrical forces, such as physical forces. Its response to an applied electric field is impeded by these forces, which have equivalent effect of increasing the relaxation time. But it is usually difficult to quantitatively relate the type of force and its magnitude to the increase in $\tau$. So, An assumption of sucrose-water mixture was assumed to develop a model for the dielectric constant of the bound-water component in the vegetation mixture [27]. Sucrose was chosen because it is a good example of the organic substances present in vegetation, and the binding arrangement of sucrose-water molecules is well known, thereby allowing them to compute the bound water concentration. Then the dielectric properties of a sucrose-water mixture can be obtained by a Cole-Cole dispersion equation [27] as

$$\varepsilon_b = 2.9 + \frac{55 + \left(1 + \sqrt{f/0.36}\right)}{\left(1 + \sqrt{f/0.36}\right)^2 + \left(f/0.36\right)} - j\frac{55\sqrt{f/0.36}}{\left(1 + \sqrt{f/0.36}\right)^2 + \left(f/0.36\right)} \tag{3}$$

Here, the factor 0.36 represents double of the relaxation frequency of bound water in GHz. This expression indicates that the dielectric constant of bound water is independent of temperature and salinity.

Substituting Eqs. (2) and (3) into Eq.(1), the evolution of complex dielectric constant of a leaf with respect to frequencies up to 2 THz is calculated and shown in Fig. 4(a). Both the real and imaginary parts decrease with the increasing of frequency, which is consistent with the measurements in [17, 19] for a fresh coffee leaf. Besides, the decrease becomes slower for frequencies above 0.5 THz, which means its frequency resolution becomes smaller at higher frequencies. In Fig. 4(b), with the increasing of water content, the dielectric constant increases significantly for THz waves at 1 THz, which indicates that the water content affects the dielectric properties of plant leaves significantly and it can be assessed by THz waves.

In order to see the influence of temperature and salinity, the variations of dielectric constant under different temperatures and salinities are plotted in Fig. 4 (c) and (d). The real and imaginary parts are almost unchanged when the temperature and salinity variations are up to 30$^\text{o}$C and 15 psu. This means the dielectric properties of plant leaves are almost immunized to the change of temperature and salinity. This is because the non-dispersive residual component and bound water is independent of both parameters as demonstrated before in Eqs. (1) and (3). The free water is very sensitive to the change of temperature and salinity in Fig. 3, but the small volume fraction of it as in Fig. 2 would not change that much. This conclusion would be further confirmed in later calculations.

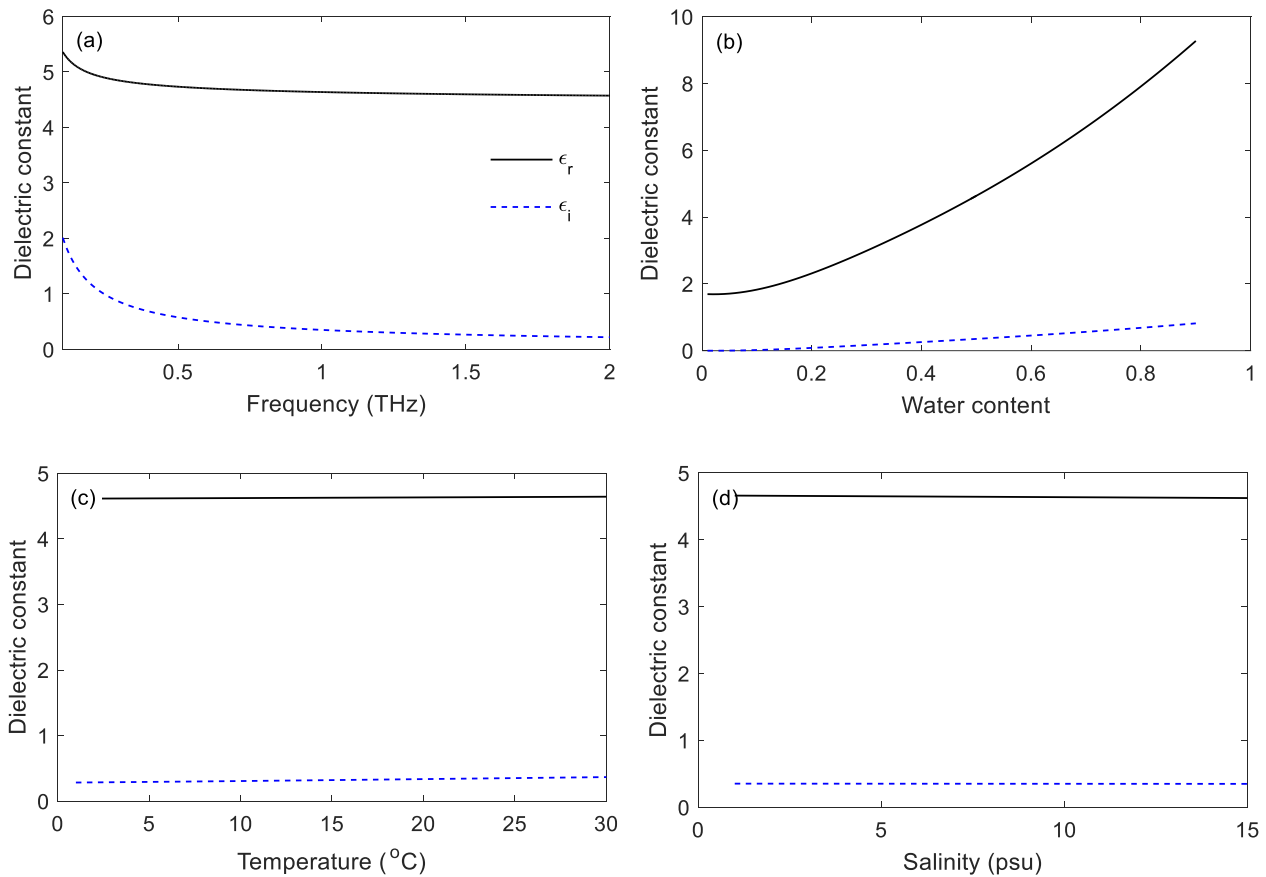

Figure 4 (a) Dielectric spectrum of a fresh leaf with water content $m_\text{g} = 0.5$, temperature $T = 25^\text{o}$C and salinity $S = 10$ psu; (b) Evolution of dielectric constant with respect to water content with frequency at 1 THz, temperature $T = 25^\text{o}$C and salinity $S = 10$ psu; Variation of dielectric constant of plant leaves with respect to (c) temperature and (d) salinity with THz frequency at 1 THz, water content $m_g = 0.5$, salinity $S= 10$ psu and temperature $T= 25^{o}$C.

## 3. Theoretical Model for Reflection Geometry

A single plant leaf can be regarded as a layered medium in parallel-plane configuration. When an EM wave encounters the medium as shown in Fig. 5(a), a portion of it would be reflected back while the rest pass through the medium. This is due to the difference in dielectric properties of the neighboring mediums, which can cause a phase shift and attenuation to the incidence wave. In addition, the transmitted wave can get trapped inside the slab and undergo multiple forward and backward reflections [19]. Such a behavior, which depends on several parameters including the thickness of the medium and the transmission frequency, can have a constructive or destructive effect on the resultant field pattern when the medium is homogeneous.

To predict the reflection by a plant leaf, there have been several methods employed by tracking all the multiple reflections that occur at the two boundaries [11, 19, 36, 37]. When using the multiple reflection methods, the total reflection coefficient can be obtained as

$$\rho = \frac{\rho_{12} + \rho_{23} \exp\left(-2 j k_2 d \cos\theta_1^{'}\right)}{1 + \rho_{12}\rho_{23} \exp\left(-2 j k_2 d \cos\theta_1^{'}\right)} \quad (4)$$

with $\rho_{12}$ as the boundary reflection coefficient between media 1 and 2, and $\rho_{23}$ as the boundary reflection coefficient between media 2 and 3. $k_2$ is the wavenumber in media 2 with a thickness of $d$. $\theta_i'$ is the propagation angle in media 2 corresponding to the incidence angle of $\theta_i$ media 1 by Snell's law. Reflectivity of the THz wave at 1 THz is plotted in Fig. 5(b). The oscillation behavior is attributed to the multiple reflections as in Fig. 5(a), which is interrelated by specific phase relationships. These relationships are associated with propagation delay between the two boundaries and the phase angles of the reflection coefficients at the two boundaries. Moreover, these phase relationships also apply to reflection by every other differential volume within the middle layer, as well as the reflection originating in the bottom medium.

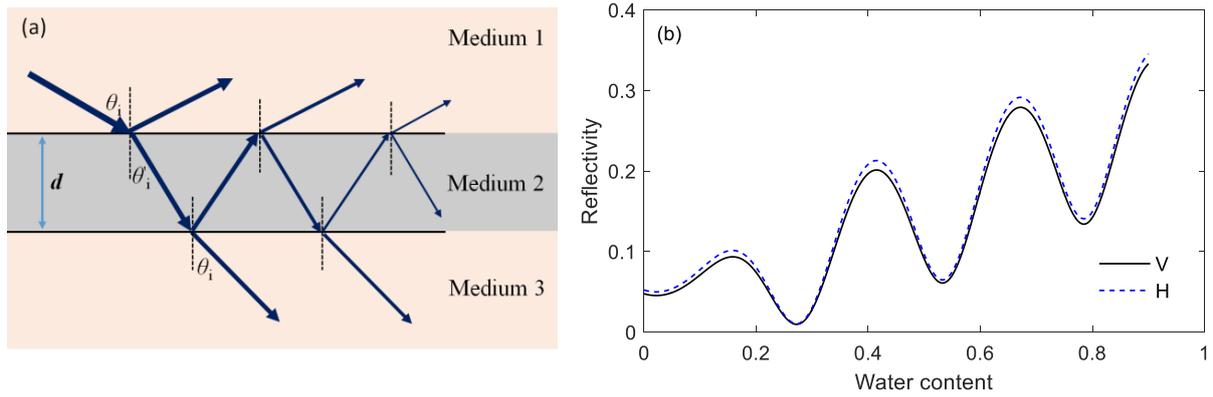

Figure 5 (a) Three-layered structure with planar boundaries; (b) Variation of reflectivity by plant leaves with respect to water content with a THz frequency at 1 THz, water content $m_g = 0.5$, salinity $S = 10$ psu, temperature $T = 25°C$ and leaf thickness $d = 0.27$mm.

In the above calculation, the leaf layer is considered to be smooth and surface roughness is not included. Here, we would regards a single leaf as a slab with an irregular top surface containing randomly distributed scatters. The integral equation method (IEM) could be employed to integrate the total bistatic scattering components over the upper hemisphere above the surface to obtain the scattering coefficient [38, 39]. This method was first developed for scattering by a conducting surface and further confirmed to be available for dielectric surfaces and medium [21]. The total scattering coefficient can be expressed as

$$\sigma_{pq}^0 = \frac{k^2}{2} \exp\left[-\sigma^2 \left(k_z^2 + k_{sz}^2\right)\right] \sum_{n=1}^{\infty} \frac{\sigma^{2n}}{n!} \left(I_{qp}^n I_{rs}^{n*}\right) W^{(n)} \qquad (5)$$

where $W^{(n)}$ is the roughness spectrum of the surface related to the surface correlation function by the Fourier transform as $W^{(n)} = \int_0^\infty \rho^n(\xi) J_0(k\xi) \xi d\xi$ with $J_0(k\xi)$ as the first spherical Bessel function and $\rho(\xi)$ as the correlation function of the surface. It can be in the form of $m$-exponential and $m$-power.

$$\rho(\xi) = \exp\left[-\left(|\xi|/l\right)^m\right] \quad m\text{-exponential} \qquad (6)$$

$$\rho(\xi) = \frac{1}{\left[1 + \left(\xi^2/l^2\right)^m\right]} \quad m\text{-power} \qquad (7)$$

The correlation functions represent different approaches to characterize the statistical correlation between the scattering height and location with another height at a location over a distance of $\xi$. The $m$-exponential correlation function is a two-parameter function that can be reduced to the exponential function for *m*=1 and Gaussian for *m*=2. The exponential function is employed in this work and the two fitting parameters could provide greater flexibility in matching the model to correlation functions based on measured surface height profiles.

The term

$$I_{\alpha\beta}^n = \left(k_{sz} + k_z\right)^n f_{\alpha\beta} \exp(-\sigma^2 k_{sz} k_z) + \frac{(k_{sz})^n F_{\alpha\beta}\left(-k_x, -k_y\right) + (k_z)^n F_{\alpha\beta}\left(-k_{sx}, -k_{sy}\right)}{2} \quad (8)$$

with $f_{\alpha\beta}$ as the Kirchhoff field coefficient and $F_{\alpha\beta}$ as the complementary field coefficient, which are correlated to the Kirchhoff and complementary fields. The parameters $k_x$, $k_y$, $k_z$, $k_{sx}$, $k_{sy}$, $k_{sz}$ are the Cartesian components of the wave vectors $k$ and $k_s$. So, the total scattering coefficient $\sigma^0_{qq}$ depends on the rms (root-mean-square) height and the surface correlation length, as well as the dielectric constant of the surface material $\varepsilon$ through the Fresnel reflectivities [39-41].

The IEM model can be used to compute the backscattering and bistatic scattering coefficient of a random surface with any specified degree of roughness for any combination of receive and transmit wave polarizations [38]. For example, in [21], the model shows good agreement with measured backscattering behavior of a rough aluminum plate at 160 GHz, 240 GHz and 1.55 THz. It has also been used to calculate the surface scattering behavior of a multilayer dielectric material [42].

When an incident angle of 10° is set to avoid standing waves (and we find that the change of reflection coefficient can be negligible when compared with normal incidence), the specular reflection by a plant leaf is calculated and shown in Fig. 6 (a) with the Gaussian correlation function considered. We can see that there is a difference between the V- and H- polarized components, even though it's below 1 dB. But this value would increase when the incidence angle is increased and would reach to 0 for normal incidence according to Fresnel's law. Another thing we should note is that the H-polarized component owns a higher reflectivity and should be employed for future detection of leaf water status in reflection geometry.

However, inhomogeneities exist inside the leaf layer and the inner volume scattering should not be neglected, even though the phase relationships among the multiply-reflected contributions are no longer preserved, in which case the reflection by the middle and bottom layer can be added as powers. For this, a general solution of the scattering problems which are the combination of the volume and surface scattering should be considered and has not been adequately developed at THz frequencies. In this part, the radiative transfer equation (RTE) would be employed with the IEM to calculate the scattering effects by the inhomogeneous vegetation-water mixture [43, 44]. A general expression for the method could be

$$\sigma_{pq}^0 = T_{12}^2(\theta_i)\left[\Upsilon^2 \sigma_{23}^0(\theta_i') + \frac{\sigma_{vpq}^{back}\cos(\theta_i')}{2\kappa_e}(1-\Upsilon^2)(1+\Gamma_{23}^2(\theta_i')\Upsilon^2) + 2\sigma_{vpq}^{bist} d\Gamma_{23}(\theta_i')\Upsilon^2\right] + \sigma_{12}^0(\theta_i) \quad (9)$$

with $T_{12}(\theta_i)$ as the boundary transmissivity between media 1 and 2, and $\Gamma_{23}(\theta_i)$ as the boundary reflectivity between media 2 and 3. $\Upsilon$ is the one-way transmissivity of the middle layer along direction $\theta_i'$, $\sigma_{12}^0(\theta_i)$ and $\sigma_{23}^0(\theta_i')$ are the backscattering coefficient of the upper and lower boundaries, which could be obtained by the IEM model. $\sigma_{vpq}^{back}$ and $\sigma_{vpq}^{bist}$ are volume backscattering and bistatic scattering coefficients, respectively. Subscripts *p* and *q* represent V- or H- polarization.

The calculation result is shown in Fig. 6(b) and different from that in Fig. 6(a) with only the IEM considered. A 2-6 dB improvement in the reflection coefficient can be observed when the reflection by the lower boundary is considered in Eq. (8), and the effect of absorption of the leaf can be counteracted due to the thin thickness of the plant leaf which is set at 0.27 mm as we measured by a micrometer screw. We can also see there is a difference between the V- and H- polarizations, and it becomes larger when the incidence angle is increasing. This indicates that H-polarization should be preferred in reflection geometries, especially for large incidence angles. Also, reflection coefficient is related to the leaf water content and can be used for leaf water status monitoring.

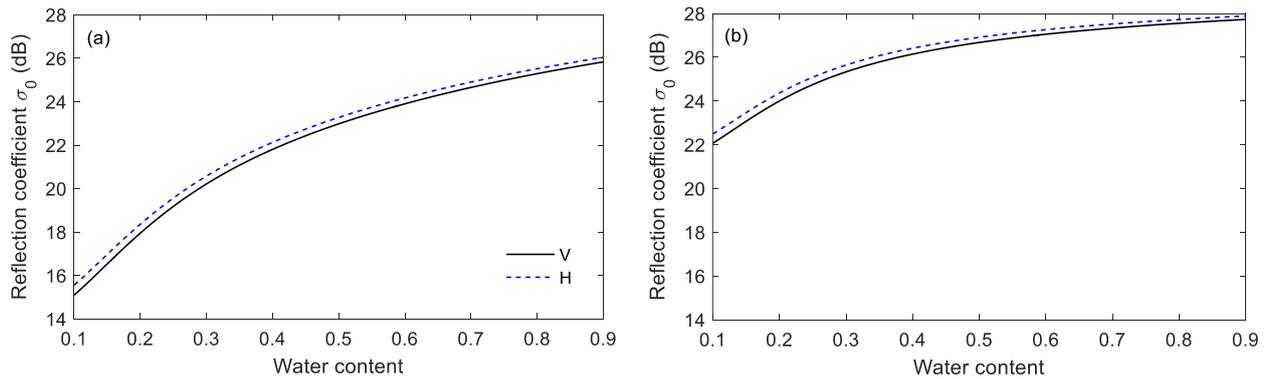

Figure 6 (a) Variation of reflection coefficient with respect to water content when (a) IEM and (b) IEM + RTE employed. Frequency $f$ = 1 THz, incidence angle $\theta_i$ = 10°, water content $m_g$ = 0.5, temperature $T$ = 25°C, salinity $S$ = 10 psu, leaf thickness $d$ = 0.27mm, upper boundary surface roughness $s_{12}$ = 10μm and correlation length $L_{12}$ = 6mm, lower boundary surface roughness $s_{23}$ = 50μm and correlation length $L_{23}$ = 6mm.

In order to see the reflection behavior at other frequencies, we calculate the reflection of H-polarized component at different frequencies as shown in Fig. 7(a). The difference of water content can be recognized over the whole frequency range, which means THz wave can be used in reflection geometry for water status monitoring. We also find that the reflection coefficient at higher frequencies is much larger than that at lower frequencies. This is consistent with the results in Fig. 5 and implies that a higher frequency should be preferred in reflection geometry. Another thing we need to note is that the difference over the same frequency variation becomes lower at higher frequencies, which means the sensitivity becomes lower. Here, in one word, we can say that the higher frequency owns higher reflection coefficient but less sensitive to water content variation when compared with lower frequencies.

To see whether the non-specular component can be used for water content monitoring, we calculate and plot the reflection coefficient with respect to water content under different scattering angles when the incidence angle is set at $10^o$. In Fig. 7(b), non-monotone curves are observed when the scattering angle is not identical to incidence angle. We attribute this to the interference of diffuse scattering components by the rough leaf surface. An identical phenomenon is observed in [38] due to diffuse scattering by a metallic rough surface. So, only the specular component can be used for water content monitoring and this is considered in the following calculation.

To see the impact of temperature and salinity on the reflection of THz waves, an evolution of reflection coefficient with respect to frequency is plotted under different temperatures in Fig. 7(c) and under different salinities in Fig. 7(d). We can see that a $30^o$C temperature variation can only lead to around a 1dB change of reflection coefficient and there is almost no change when the salinity is changed from 1 psu to 15 psu. This indicates that this method employing THz wave for monitoring is not sensitive to the change of temperature and salinity, which is further confirmed in [45].

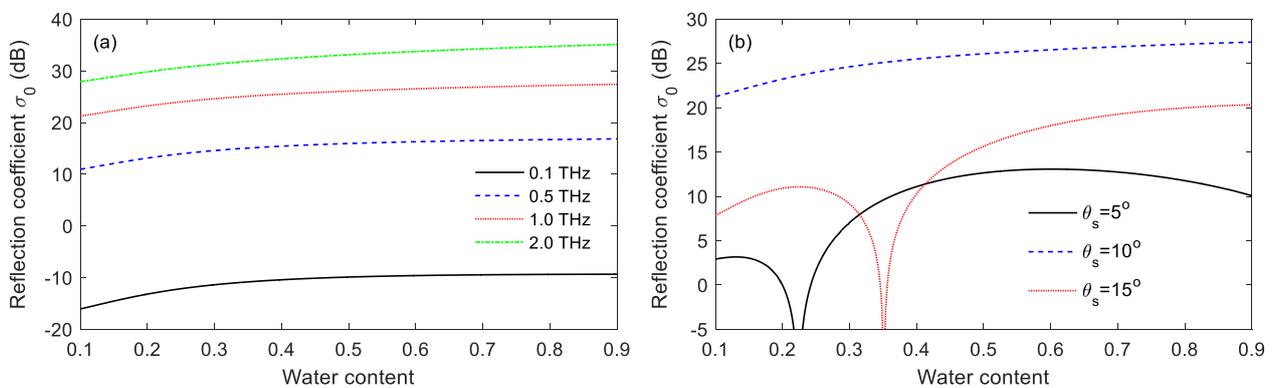

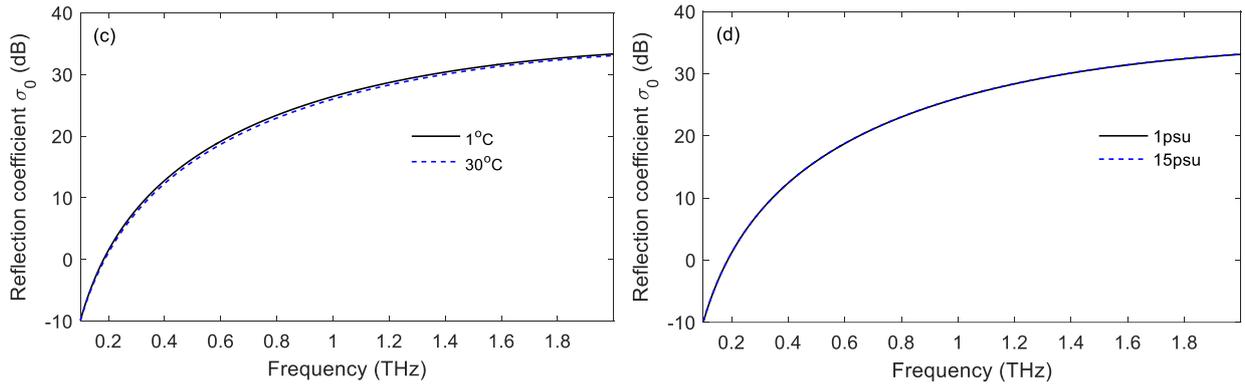

Figure 7 Variation of reflection coefficient different (a) frequencies, (b) scattering angles, (c) temperatures and (d) salinities. Incidence angle $\theta_i = 10^o$, water content $m_g = 0.5$, temperature $T = 25^oC$, salinity $S = 10$ psu, leaf thickness $d = 0.27$mm, upper boundary surface roughness $s_{12} = 10\mu m$ and correlation length $L_{12} = 6$mm, lower boundary surface roughness $s_{23} = 50\mu m$ and correlation length $L_{23} = 6$mm.

## 4. Theoretical Model for Transmission Geometry

In transmission geometry, a coherent transmission theory employing the propagation matrix is usually employed, where a leaf is regarded as a homogeneous layer [19]. But here, we would consider the incoherent transmission which occurs when the leaf layer has randomly distributed inhomogeneities with dimensions are around and larger than $\lambda/100$. The phase coherence among the multiply-reflected contributions is no longer preserved, and the oscillation behavior observed in Fig. 5(b) disappears . From the rediative transfer theory [31], we can obtain the transmissivity as

$$T = \left(\frac{1-\Gamma_{12}}{1-\Gamma_{12}\Gamma_{23}\Upsilon^2}\right)\left[(1+\Gamma_{23}\Upsilon)(1-a)(1-\Upsilon)+(1-\Gamma_{23})\Upsilon\right] \qquad (10)$$

with $a$ as the single scattering albedo of the middle layer.

Fig. 8 shows the transmissivity of THz waves through the plant leaf layer. We can see that the V-polarized component owns higher transmission when the incidence angle is $10^o$, and this difference becomes more obvious for larger incidence angles. This is consistent with the result in Fig. 5, where the H-polarizated component owns a higher reflection. So the V-polarization should be employed for latter calculation in transmission geometry. Another thing we can see is that, opposite to the reflection geometry, lower frequencies owns higher transmission here. In Fig. 8(b), the transmission becomes lower when water content $m_g$ increases. Also, in Fig. 8(c) and (d), we can find that the change of temperature and salinity on the transmission variation is smaller than 1 dB and is negligible.

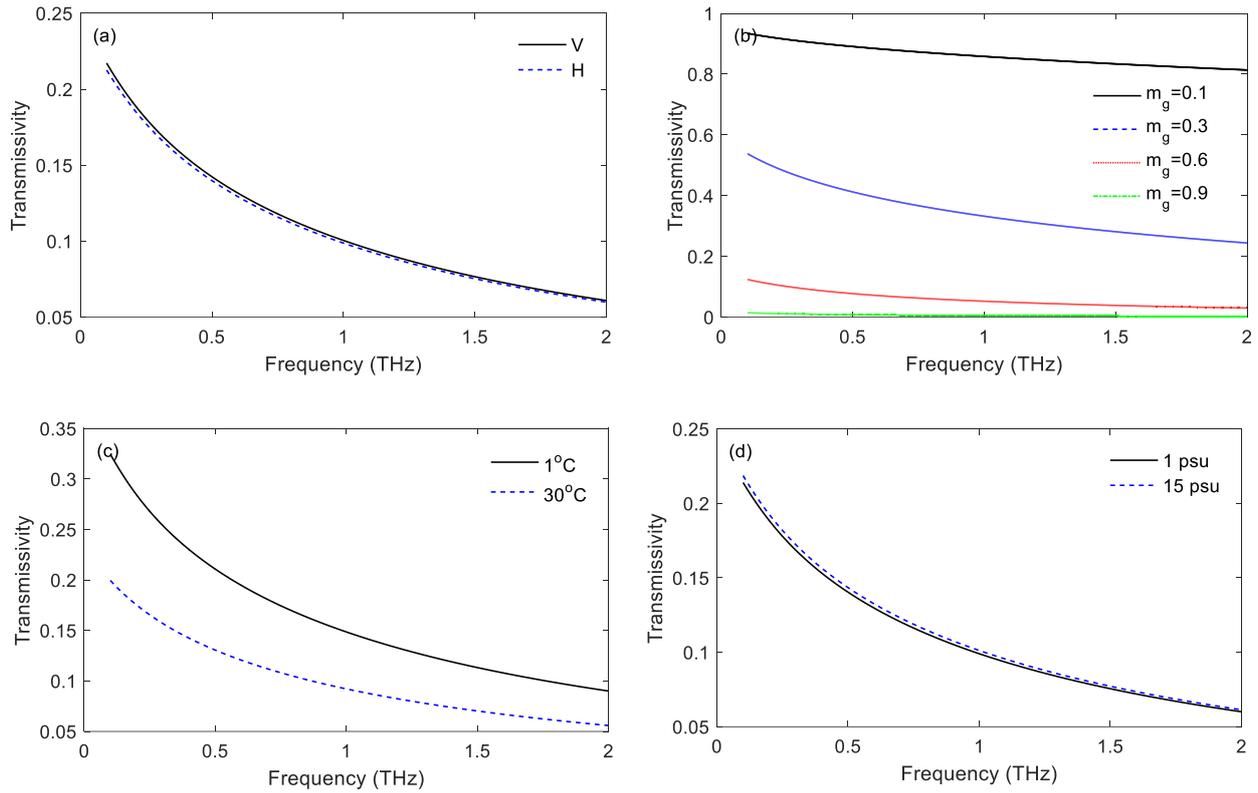

Figure 8 Variation of THz transmissivity with respect to frequency under different (a) polarization, (b) water content, (c) temperature and (d) salinity. Incidence angle $\theta_i = 10^o$, water content $m_g = 0.5$, temperature $T = 25^oC$, salinity $S = 10$ psu, leaf thickness $d = 0.27$ mm.

To verify the validity of our method in leaf water status prediction, a standard commercial THz time-domain spectroscopy system (TERA K15 by Menlo Systems Company, Germany) is employed and operating in transmission geometry [46] for measurements. The transmitter and receiver are coupled with a collinear adapter to perform as a collinear transmission transceiver. A rapid scanning rate of 20 Hz, and a focal spot of 2 mm in diameter could be employed and the raster scan minimum step size is 2.5 mm and the imaging area of the XY-stage is up to $5 \times 5$ cm$^2$. In the detection process, the substrate as placed at the focal point of the detector to test the reference signal in destruction method and air is tested as the reference in non-destruction method, followed by the sample signal detection. Fresh leaves of winter wheat (Jingdong 24) were cultivated in Beijing Research Center for Information and Technology in Agriculture for samples, which are placed at the focal point of the detector to test the transmitted signal. The fresh leaves were first used for THz spectrum measurements, and then weighed by an accurate electronic balance. Each of these leaves was tested individually and repeatedly every time at the same conditions. All the leaves were kept in self-sealing bags after they were taken from plants and then placed in an incubator (4$^o$C) to decrease the water evaporation and tested immediately. Before the measurement, the leaves were wiped by sterile cotton, and then fixed onto the sample holder after the thickness was measured. The vegetation gravimetric water content of winter wheat ($m_g$) in nature can change in the range of 0.5-0.85 [47], but here we can change that from 0.1 to 0.6 by a drier. We obtain the measured data at 3 different points on the middle of the leaf as the inset in Fig. 8 and repeated 2 times to confirm it can be reproduced. The measurements were conducted on 3 different leaves and the averaged transmissivity with error bar was plotted in Fig. 9. A discrepancy is observed between the measurement and calculation. We attribute this to fixed scattering parameters and leaf thickness in the model, which should change during the hydration process actually. But the evolution is not obtained in this work due to the lack of tools with high precisions.

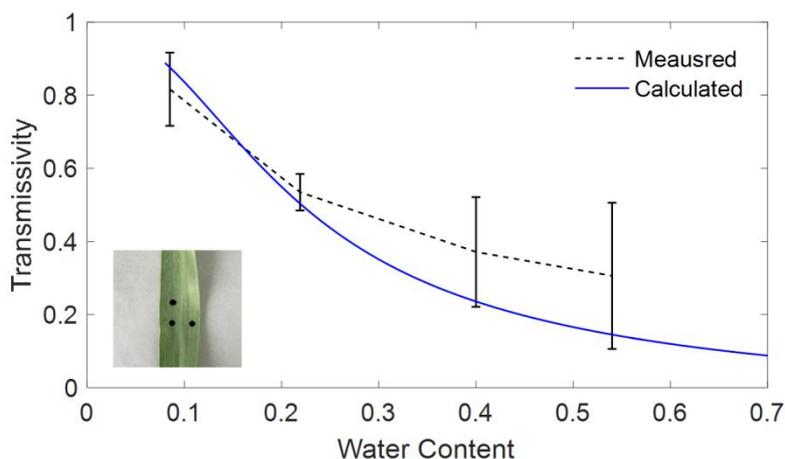

Figure 9 Measured and calculated THz transmissivity versus water content with incidence angle $\theta_i$ = 0, temperature $T$ = 25°C, salinity $S$ = 10 psu and leaf thickness $d$ = 0.27mm. The vertical error bars with 95 % confidence interval shown in the plot represent the standard deviation of recorded data. The insert shows a leaf sample we used with three black dots representing the measured positions.

**Conclusions**

In this paper, we investigate the scattering behavior of THz waves when reflected by and transmitting through a plant leaf under different water content. A theoretical model combining IEM and RTE models are firstly presented to investigate the interaction of THz wave and plant leaves. V- and H-polarized waves are more preferred for water status evaluation in transmission and reflection geometries, respectively. The influence of temperature and salinity on the scattering performance is negligible in both geometries. The theoretical prediction is in rough qualitative agreement with the measured data when the variation of scattering parameters and leaf thickness is not considered during the hydration process.

## Acknowledgements

This research was supported by Beijing Institute of Technology Research Fund Program for Young Scholars (No. 3050011181910), the Science and Technology Innovation Special Construction Funded Program of Beijing Academy of Agriculture and Forestry Sciences (No. KJCX20180119), National Key Research and Development Program of China (No. 2016YFD0702002) and Beijing Municipal Natural Science Foundation (No. 6182012).